\documentclass[11pt,twoside]{article}
\usepackage{asp2004}
\usepackage{psfig}
\usepackage{epsf}
\usepackage{graphics}
\usepackage{lscape}
\begin{document}
\title{DENIS detections of highly obscured galaxies in the area around
 PKS\,1343\,--\,601
 }
\author{Anja C. Schr\"oder}
\affil{Dept.\ of Physics \& Astronomy, University of Leicester, University
 Road, Leicester LE1 7RH, UK}
\author{Ren\'ee C. Kraan-Korteweg }
\affil{Depto. de Astronom\'{\i}a, Universidad de Guanajuato,
 Apdo. Postal 144, Guanajuato GTO 36000, Mexico }
\author{Gary A. Mamon }
\affil{IAP (CNRS UMR 7095), 98~bis Blvd Arago, 75014 Paris, France, and \\
 GEPI (CNRS UMR 8111), Observatoire de Paris, 92195 Meudon, France}
\author{Patrick A. Woudt}
\affil{Dept.\ of Astronomy, University of Cape Town, Rondebosch 7700, South Africa}

\begin{abstract}
We present results of a search for galaxies around the highly obscured
giant radio galaxy PKS\,1343\,--\,601 using the near-infrared DENIS
survey. We compare our findings with surveys in the $B$-band, at the
21\,cm line emission, and with 2MASS. Recession velocities of galaxies
in this area suggest a low-velocity-dispersion group or cluster of galaxies
including an X-ray confirmed Seyfert 2. The colours of the galaxies
have been used to examine the extinctions in this low-latitude area
where IRAS/DIRBE estimates are unreliable. We find the true extinction
to be roughly 15\% lower than the IRAS/DIRBE extinctions.
\end{abstract}
\thispagestyle{plain}

\section{Introduction}

PKS\,1343\,--\,601 (Centaurus B) is a very strong radio source in the
southern Zone of Avoidance (ZoA). Near-infrared (NIR) observations
revealed it to be a giant elliptical galaxy as often found in
the centre of galaxy clusters. At a distance of $\sim\!10\deg$ from the
Great Attractor (GA) and with a recession velocity of 3872\,km\,s$^{-1}$ (West
\& Tarenghi 1989) it lies within the GA overdensity. Finding a new
cluster here will have considerable impact on the local velocity
field calculations. On the other hand, X-ray
observations with ASCA have only revealed diffuse
emission from PKS\,1343\,--\,601 itself (Tashiro et~al.\ 1998, see
also the discussion in Ebeling et~al.\ 2002) which would rule out a rich
cluster. A more sensitive XMM-Newton observation does not indicate any
obvious cluster emission either but will have to be analysed in detail for an
improved estimation.

A detailed discussion of PKS\,1343\,--\,601 and the potential for a
cluster in this area can be found in Kraan-Korteweg et~al.\ (these
proceedings), who present $I$-band observations of a $2\deg
\times 2\deg$ area around PKS\,1343\,--\,601. Nagayama et~al.\ (2004)
have analysed deep $H J K$ observations of the inner $36\arcmin \times
36\arcmin$ (see also Nagayama et~al., these proceedings). A more
detailed presentation of the following work will be given in
Schr\"oder et~al.\ (2004). 

To search for galaxies in this highly obscured area ($\left\langle A_B\right \rangle
\simeq 12^{\rm m}$) we have used the NIR survey DENIS (Epchtein 1997,
1998). The NIR has the advantage that it is less affected by
the foreground extinction as compared to the optical (the extinction
in $K$ is about 10\% of the extinction in $B$), and that it is sensitive to
early-type galaxies, tracers of massive groups and clusters which are
neither uncovered in far infrared surveys nor in the 21\,cm radiation
where extinction is negligible (see Mamon 1994).

\section{The DENIS survey } \label{denis}

The DENIS survey is a NIR survey of the southern sky in the
Cousins-$I\,(0.8\,\mu m)$, $J\,(1.25\,\mu m)$ and $K (2.15\,\mu m)$
passbands with a resolution of $1\arcsec$ in $I$ and $3\arcsec$ in
$J$ and $K$ (Epchtein 1997, 1998).  The observations were carried out
between 1996 and 2001 with a dedicated 1\,m ESO telescope at La Silla
(Chile).  About 92\% of the southern sky ($+2\deg \le {\rm DEC} \le
-88\deg$) were covered.

DENIS images are $12\arcmin \times 12\arcmin$ large with an exposure time
of 9\,s. The observing mode consisted of step-and-stare scans of 180 images
in declination, resulting in strips of $12\arcmin \times 30\deg$. There is
an overlap region of $1\arcmin$ on each side of every image.  The limiting
magnitudes (at a sensitivity of about $3\sigma$) are $18\fm5$, $16\fm5$,
$13\fm5$ for the $I$-, $J$-, and $K$-bands, respectively, while the limits
for high completeness and reliability of the galaxy extraction at high
galactic latitudes are $16\fm5$, $14\fm8$, and $12\fm0$, respectively
(Mamon 1998).


Interpolating from Cardelli et~al.\ (1989), the extinctions in the DENIS
NIR passbands are $A_I\!=\!0\fm45$, $A_J\!=\!0\fm21$, and
$A_{K}\!=\!0\fm09$ for $A_B\!=\!1\fm0$. Thus the decrease in number
counts of galaxies in the ZoA as a function of extinction is
considerably slower in the NIR than in the optical (Schr\"oder et~al.\ 
2004, Schr\"oder, Kraan-Korteweg, \& Mamon, 1999). While DENIS easily
detects galaxies up to extinctions of $A_B \simeq 10^{\rm m}$, we
can find intrinsically bright and close by galaxies in the $K$-band
at much higher extinctions.

\section{DENIS detections }

We have searched 29 DENIS slots each with 37 images around
PKS\,1343\,--\,601. The total area covered amounts to $\sim 29.8$ square
degrees. Using the DENIS visualisation package {\tt Denis3d} by E. Copet
 we scanned each image simultaneously in the three passbands. That way
the relative appearance of highly-obscured galaxies in the three bands,
which is different to most stars, could be used to distinguish galaxies
from blended stars and extended Galactic objects. 

The automatic extraction package {\tt SExtractor} (Bertin \& Arnouts
1996) was used to obtain $I J K$ photometry for the visually detected
galaxies. We have derived total magnitudes as well as colours within a
7-arcsecond aperture (to avoid strong contamination by superimposed
stars). We have subsequently checked the magnitudes and colours of
each galaxy and compared them with the image to see whether {\tt
SExtractor} has deblended all objects close by and whether all the
parameters agree with each other.

The Galactic foreground extinctions has been determined using the
IRAS/ DIRBE maps by Schlegel, Finkbeiner, \& Davis (1998): The colour
excess $E_{B-V}$ has been converted to $A_B$ using $R_B=4.14$ (Cardelli
et~al.\ 1989). Note that measurements at latitudes $|b|<5\deg$
have not been properly calibrated and are thus uncertain.

\begin{figure}[tb]
\plotfiddle{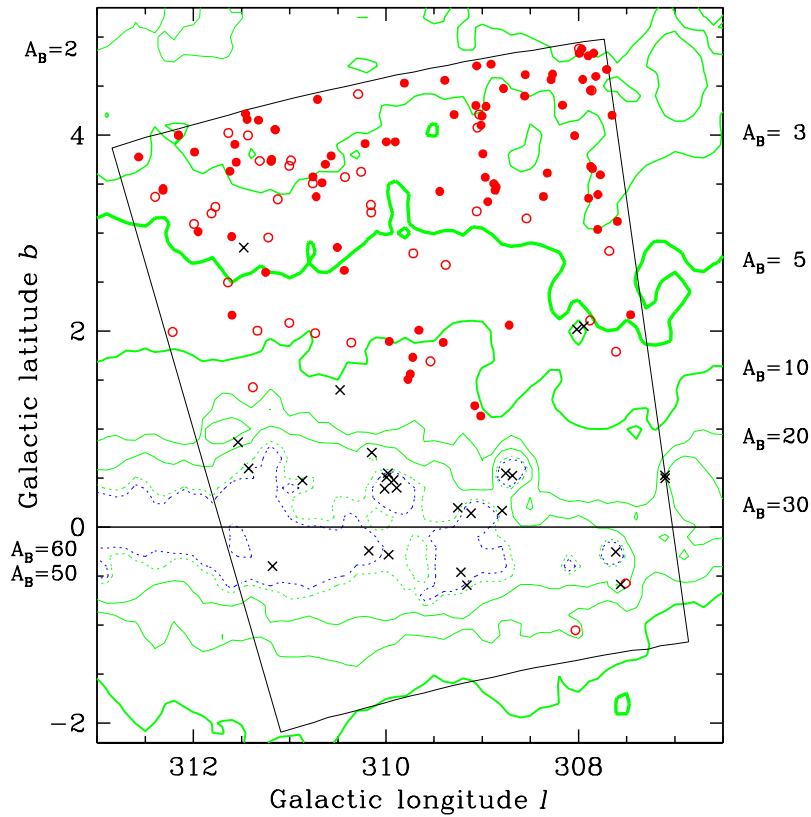}{10.2cm}{0}{55}{55}{-170}{-55}\\
\vspace{0.8cm}
\caption[]{The distribution of galaxies (circles) and Galactic objects
(crosses) in the searched area (tilted rectangular) is shown.  Filled
circles are galaxies visible with DENIS, and open circles stand for
uncertain DENIS galaxies. Extinction contours according to the
IRAS/DIRBE maps are displayed as labelled. The galaxy
PKS\,1343\,--\,601 ($l=309\fdg7, b=+1\fdg8$, $A_B=12\fm3$) is close to
the centre in this image.
  } 
\label{clnplot}
\end{figure}

On 1073 searched images we found 83 galaxies (plus 38 uncertain candidates). Of
these, 79 (33) are visible in the $I$-band, 82 (35) in $J$, and 67
(25) in $K$ (see Schr\"oder et~al.\ 2004 for details). 
Figure~\ref{clnplot} shows the distribution of the detected galaxies:
filled circles stand for definite galaxies, and open circles show
uncertain candidates. Crosses denote Galactic objects. The contours
depict the extinctions at $A_B=2^{\rm m}$, $3^{\rm m}$, $5^{\rm m}$,
$10^{\rm m}$, $20^{\rm m}$, $30^{\rm m}$, $50^{\rm m}$, and $60^{\rm
m}$, as indicated.

While PKS\,1343\,--\,601 is prominent in all three passbands, a nearby
equally large galaxy is not visible in the $I$-band and is probably
an early-type spiral with an elongated halo. Two objects at very high
extinctions (with $A_B = 39^{\rm m}$ and $24^{\rm m}$, for DZOA4641-06
and DZOA4645-13, respectively) are classified as uncertain galaxies;
they are only visible in the $K$-band. With extinction-corrected (but
not diameter-corrected, {\it cf.}\ Cameron 1990) magnitudes of $K^o = 8\fm8$
and $8\fm3$, respectively, they are only slightly fainter than the
central elliptical galaxy at $K^o = 8\fm0$, which would indicate that
these two galaxy candidates could lie at the distance of the cluster
or closer.

At very high extinctions, we also detected extended Galactic objects
like planetary nebulae, \protect\normalsize H\thinspace\protect\footnotesize
II\protect\normalsize\ regions, and reflection nebulae, which
are often associated with young stellar objects. Some of these objects
are easily recognisable as non galaxian, but there are cases where a
distinction could only be made on the basis of high foreground
extinction (which is usually $A_B > 50^{\rm m}$). A careful
comparison between unambiguous galaxies and Galactic objects was used
to classify cases at intermediate extinctions.

\section{Comparison with other catalogues } \label{lit}

\subsection{$B$-band} \label{wkk}

Woudt \& Kraan-Korteweg (2001, hereafter WKK), who have conducted a
deep search for galaxies using copies of the ESO-SERC $B_J$-band photographic
plates, list 35 galaxies within our search area. Only one of their galaxies (WKK2589,
DZOA4655-08) is not visible on the DENIS images; it is a small, very
low surface brightness galaxy, probably of type Sm or Irr. Five $B$-band
galaxies, all classified as uncertain galaxies by WKK, were identified
as (blended) stars with the higher spatial resolution of the DENIS $I$-band images.

Most $B$-band galaxies are found in the low extinction regions, that is, at
$A_B<3^{\rm m}$. The highest extinction for a $B$-band galaxy in the
searched area is $A_B=5\fm2$ (DZOA4641-07, WKK2301). The completeness
limit for $B^o=15\fm5$ and $D^o=60\arcsec$ of the WKK-catalogue is
$A_B=3^{\rm m}$.

\subsection{2MASS} \label{2mass}

We have extracted 65 objects from the 2MASS all-sky extended source
catalogue (2MASS, 2003) within the region $13\fh45<$\,RA\,$<14\fh11$ and
$-63\fdg8<$\,Dec\,$<-57\fdg4$, 11 of which are just outside the searched
DENIS area. Forty nine objects are common in both data sets, nine of which
are classified by us as Galactic objects and two as uncertain galaxies.
Five 2MASS objects were not found with DENIS; one is a Galactic object, two
are mis-identifications, and two are very small but bright galaxies that
were not recognised as such in the DENIS search. This gives a rough
estimate of the reliability of 80\% for the automatic 2MASS extraction in this highly
confused and obscured region, while the completeness of the 2MASS galaxy
extraction in this area is $\sim\!50$\%.
Many of the DENIS galaxies which are not in the 2MASS extended source
catalogue were either not recognised as extended by the 2MASS automatic
search algorithm (they will be in the 2MASS point source catalogue however), or
they are very faint in $J$ and $K$ and were only found through the
DENIS $I$-band. 

\begin{figure}[tb]
\plotfiddle{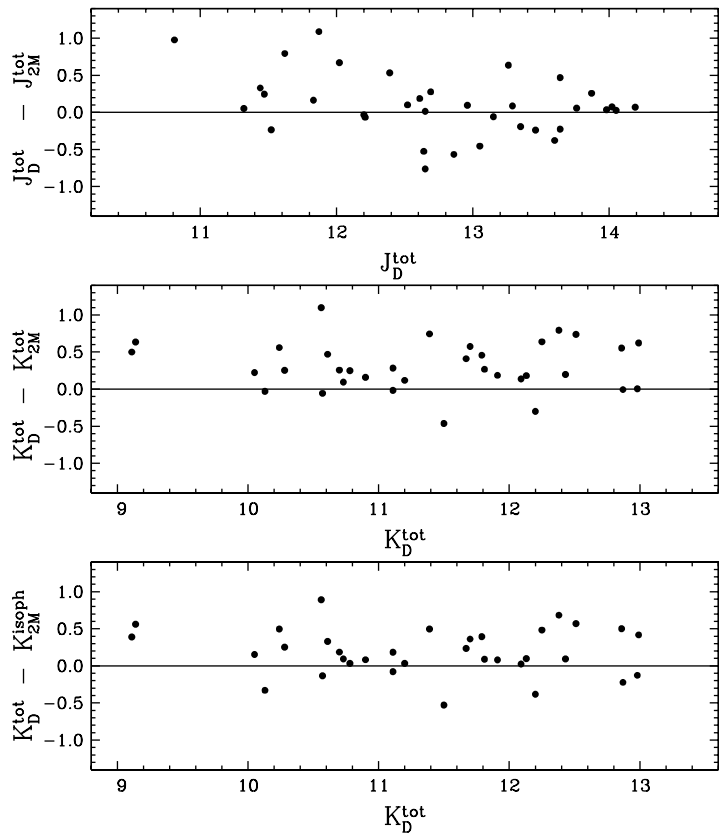}{10.5cm}{0}{50}{50}{-160}{-35}\\
\vspace{0.5cm}
\caption[]{The difference of DENIS and 2MASS magnitudes versus
DENIS magnitudes in $J$ (upper panel), in $K$ (middle panel) and for the
2MASS isophotal $K$-band magnitude in the bottom panel. 
  } 
\label{d2mplot}
\end{figure}

Figure~\ref{d2mplot} shows a comparison of the 2MASS $J$- and $K$-band
magnitudes with our magnitudes.  The $J$-band shows a magnitude difference
of $0\fm08\pm0\fm07$ and a standard deviation of $0\fm41$ (upper panel),
while in the $K$-band (middle panel) the differences is $0\fm30\pm0\fm06$
with a standard deviation of $0\fm33$, that is, there is a substantial
offset between the two catalogues in the $K$-band. This offset is due to
the large difference in magnitude limits of the two surveys: the
high-latitude ($|b| > 20\deg$) nominal values for 2MASS are $14\fm3$ and
$15\fm8$ for $K$ and $J$, respectively, while for DENIS they are $12\fm0$
and $14\fm8$, respectively. If we use the 2MASS isophotal $K$-band
magnitudes (at $20^{\rm m}$ per square arcsecond) instead the mean of the
differences becomes smaller at $0\fm18\pm0\fm06$ ($\sigma = 0\fm32$, bottom
panel), indicating that the DENIS isophotal limit out to which {\tt
SExtractor} determines the total $K$-band magnitudes is even brighter.

Using the 2MASS isophotal $J$-band the change in offset is negligible
(the mean is now $0\fm03\pm0\fm07$ with $\sigma = 0\fm34$; not shown)
as expected by the smaller difference in limiting magnitudes.

\subsection{X-rays}

One of the galaxies found in our DENIS search, DZOA4653-11, was
discovered in the serendipitous 1XMM catalogue (XMM SSC, 2003) and was
identified as a highly obscured AGN (Motch, priv.\ comm.)  with high
foreground extinction. A deep $I$-band image obtained by Motch
indicates an early type spiral galaxy.  A prominent H$\alpha$ line in
the optical spectrum obtained with EFOSC at the 3.6\,m ESO telescope
by Motch leads to a heliocentric velocity of $v=3643$\,km\,s$^{-1}$, which
means this galaxy is a cluster member. The X-ray flux over the energy
range $0.2 - 12.0$\,keV is slightly variable with time and gives
$(1.351\pm 0.007)\times 10^{-11}$\,erg\,cm$^{-2}$\,s$^{-1}$, {\it i.e.},\ 
$L_X = 2.2 \times 10^{42}
h_{70}^{-2} \rm erg\,s^{-1}$.

\begin{figure}[tb]
\plotfiddle{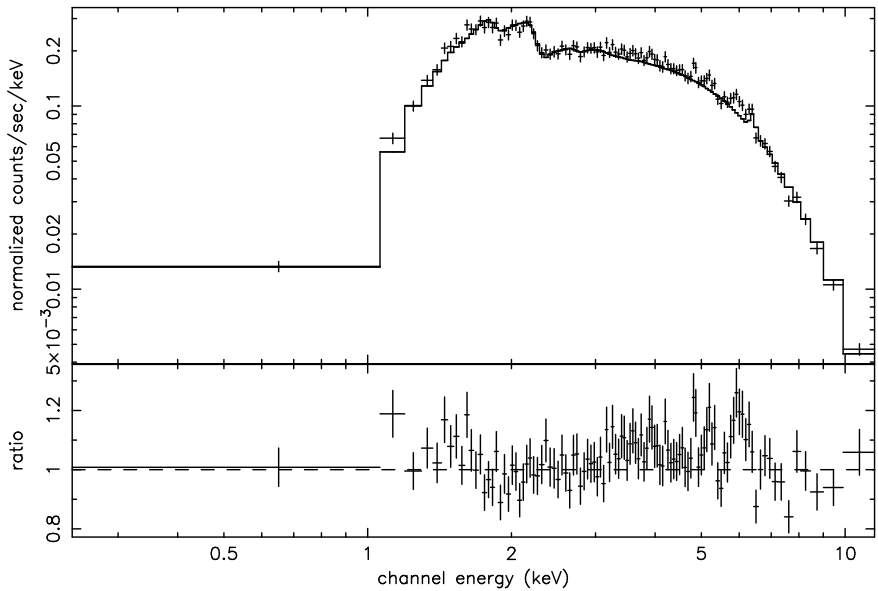}{9.5cm}{270}{50}{50}{-190}{335}\\
\vspace{-2.cm}
\caption[]{The X-ray spectrum of the AGN DZOA4653-11 with a model of
a cold absorber plus a power law fitted, top panel, and the residuals
of the fit, bottom panel.
  } 
\label{xmmplot}
\end{figure}

Fitting a cold absorber plus a power law to the XMM-Newton spectrum
(Fig.~\ref{xmmplot}) we find a photon index of $1.27\pm0.02$ and a
hydrogen column density of $N_H = (2.05\pm 0.04)\times
10^{22}$\,cm$^{-2}$. The Galactic column density at this position from
the IRAS/DIRBE estimate from Schlegel et~al.\ is only $N_H = 1.07\times
10^{22}$\,cm$^{-2}$, indicating there is significant intrinsic
absorption (note, though, that the IRAS/DIRBE maps are uncertain at
this latitude), suggesting it to be a Seyfert 2 galaxy. A faint iron line
(44\,eV equivalent width) is also visible.

\subsection{21\,cm wavelength } \label{hi}

At 21\,cm, the Galactic plane is transparent and the
search for galaxies here is limited because of the increased noise due
to strong Galactic continuum sources. We have extracted 
\protect\normalsize H\thinspace\protect\footnotesize
I\protect\normalsize\ 
velocities for galaxies in this area from the blind \protect\normalsize H\thinspace\protect\footnotesize
I\protect\normalsize\ Parkes All
Sky Survey ZoA (HIZOA) survey of the southern ZoA conducted with the
multibeam receiver on the Parkes telescope (Kraan-Korteweg et~al.\ 
2004), which covers the entire southern ZoA ($|b| \leq 5\deg$) in the
velocity range $-1200$ to 12\,700\,km\,s$^{-1}$ with an integration time of 25
minutes.  To these, we added 7 optical velocities of galaxies within
the search area (Fairall et~al.\ 1998; Visvanathan \& Yamada 1996; West
\& Tarenghi 1989; and the AGN DZOA4653-11).

\begin{figure}[tb]
\plotfiddle{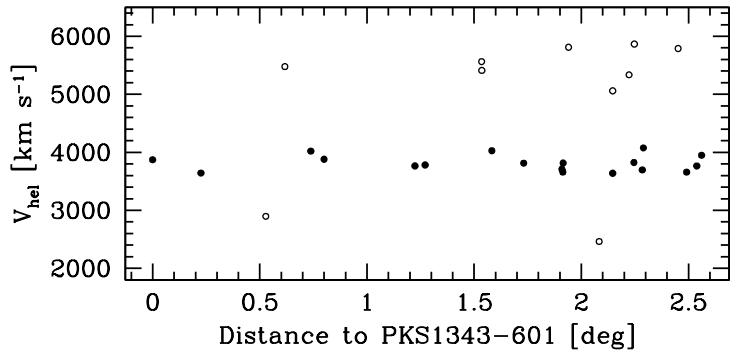}{4.5cm}{0}{50}{50}{-160}{-190}\\
\caption[]{Radial velocities as a function of distance to PKS\,1343\,--\,601. The
filled circles represent the low velocity dispersion sub-system while
the open circles show the velocity outliers.
  } 
\label{velplot}
\end{figure}

Figure~\ref{velplot} shows the velocities plotted as a function of the
distance to the giant elliptical PKS\,1343\,--\,601. The velocities in close multiplets and pairs
were replaced by their median values, so as to avoid double counting multiply-identified
single objects, as well as to remove the binary contribution to the group
velocity dispersion. The velocity
dispersion out to $2\deg$ and for $1850 < v < 6500\,\rm km\,s^{-1}$ is
$\,907\,\rm km\,s^{-1}$. However, Figure~\ref{velplot} shows wide gaps in
the velocity distribution. In the range $3000 < v < 4700\,\rm km\,s^{-1}$
there are 18 galaxies with a velocity dispersion $\sigma_v = 136\,\rm
km\,s^{-1}$, and the gaps outside of the extreme velocities among
these 18 galaxies correspond to over $5\,\sigma_v$.  Taking the velocity
dispersion at face value, we estimate the virial radius of the system to be
$233 \, h_{70}^{-1} \, \rm kpc = 0\fdg5$, but we only have two velocities
within this distance to PKS\,1343\,--\,601, and, iterating over assumed
values of the virial radius, one finds that the velocities within the
virial radius are compatible with the idea that the system around
PKS\,1343\,--\,601 is richer with a larger true velocity dispersion,
although it cannot be as rich as a rich cluster (see Schr\"oder et~al.\
2004, for details). This is consistent with the upper limit on the cluster
X-ray emission (Ebeling et~al.\ 2002). However, more velocity measurements
and deeper NIR imaging will be needed to confirm this result.


\section{Extinction and NIR colours }  \label{disc}

Colours of galaxies are independent of distance but sensitive to
extinction. Hence, NIR colours can be used to derive extinctions in
low latitude areas where the IRAS/DIRBE maps are not properly
calibrated. Cameron (1990) has shown that one needs to apply both a
correction to the isophotal magnitude of a galaxy as well as for the
reduction of the major axis of an obscured galaxy.  Since the colours
used for the following analysis were derived from 7-arcsecond
apertures (where contamination by superimposed stars is small) we only
applied the magnitude correction according to the extinction in the
IRAS/DIRBE maps and looked for dependencies in the residuals.

\begin{figure}[tb]
\plotfiddle{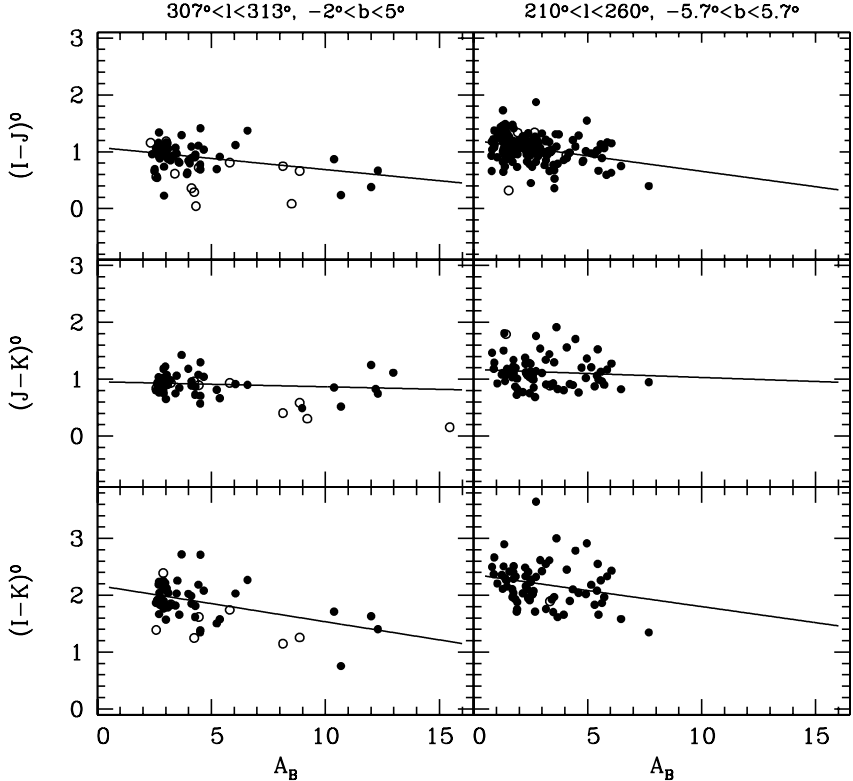}{10cm}{0}{60}{60}{-200}{-100}\\
\vspace{-0.cm}
\caption[]{The NIR colours, corrected for extinction according to the
IRAS/DIRBE maps, are plotted versus extinction in the $B$-band for the
galaxies around PKS\,1343\,--\,601 (left panels) and for galaxies around
\protect\small H\thinspace\protect\scriptsize
I\protect\small\ detections in a different region of the ZoA (right panels). Filled
circles are galaxies, open circles are uncertain galaxies. Least square
fits to the filled circles are shown. 
  } 
\label{colextplot}
\end{figure}

The left-hand panels in Fig.~\ref{colextplot} show the
extinction-corrected colours as a function of extinction in the $B$
-band (filled circles are galaxies, open circles uncertain
candidates). Galaxies at higher extinctions are clearly too blue, {\it i.e.},\ 
the extinction correction according to the IRAS/DIRBE maps
overestimates the true extinction.

With the least square fits to the filled circles ({\it i.e.},\ excluding the
uncertain candidates), written in the form $C^o = a A_B + b$ where $C$
is the colour, we find:
\[(I-J)^o = (-0.038 \pm 0.013) \,A_B + (1.071 \pm 0.057), \ \ \sigma = 0.21 \]
\[(J-K)^o = (-0.008 \pm 0.009) \,A_B + (0.940 \pm 0.049), \ \ \sigma = 0.18 \]
\[(I-K)^o = (-0.061 \pm 0.018) \,A_B + (2.149 \pm 0.086), \ \ \sigma = 0.28 \]
If we assume the true extinction to be a constant factor $f$ of the
IRAS/DIRBE value of Schlegel et al., {\it i.e.},\ $A_B^{\rm true} = f\,A_B$, we can
derive the correction from the above equations as 
\[f = 1 + {a \over E/A_B},\]
where $E$ is the reddening in the respective colour. Combining the
(independent) estimates of $f$ for \mbox{$I-J$} and \mbox{$J-K$}, we derive
\[A_B^{\rm true} = (0.86 \pm 0.04)\,A_B.\]

To verify this result and to exclude the possibility that this
particular region shows an unusual deviation, we have also derived
DENIS colours of galaxies found in a different region of the
ZoA. These are part of a project that looks for DENIS counterparts of
galaxies found with the blind \protect\normalsize H\thinspace\protect\footnotesize
I\protect\normalsize\ ZoA shallow survey (HIZSS, Henning
et~al.\ 2002) within a 10-arcminute
search radius. Forty \protect\normalsize H\thinspace\protect\footnotesize
I\protect\normalsize\ detections in the area $210\deg < l < 260\deg$, $|b| <
5\fdg7$ have been searched so far (Schr\"oder et al., in preparation).
The right-hand panels in Fig.~\ref{colextplot} show the results for all
galaxies found in these regions and with good photometry. The least square fits to the
filled circles give:
\[(I-J)^o = (-0.054 \pm 0.013) \,A_B + (1.199 \pm 0.037), \ \ \sigma = 0.22 \]
\[(J-K)^o = (-0.014 \pm 0.019) \,A_B + (1.171 \pm 0.065), \ \ \sigma = 0.26 \]
\[(I-K)^o = (-0.056 \pm 0.027) \,A_B + (2.365 \pm 0.092), \ \ \sigma = 0.36 \]

Despite the different coverage in extinction the slopes of the least
square fits agree well within the errors.  Combining now the estimates
for both regions in the colours \mbox{$I-J$} and \mbox{$J-K$} we derive
\[A_B^{\rm true} = (0.83 \pm 0.03)\,A_B.\]
In other words, we find the true extinction in the southern ZoA to be
about 17\% lower than the estimates of the maps by Schlegel
et~al.\ (1998). 

One has to keep in mind that, apart from the difficulties in interpreting
high IRAS/DIRBE measurements as reddening values, the gas/dust
distribution in the Galactic Plane shows variations on a small spatial
scale. This patchiness may lead to a bias in finding galaxies
preferentially in the low extinction areas and therefore to
underestimate the true extinction. However, the agreement between the
two regions we have used is excellent despite the fact that one covers
a range in extinction of $2^{\rm m}<A_B<16^{\rm m}$ while the other is
restricted to $0^{\rm m}<A_B<8^{\rm m}$. In other words, the fraction
of galaxies missed in the range $8^{\rm m}<A_B<16^{\rm m}$ due to much
higher extinction patches is negligible and does not affect our result.

We will continue with this work using the colours of galaxies found
within the search radii of the other half of the 110 HIZSS detections
at higher Galactic longitudes (and often at higher extinctions) to
improve the statistics. Future deeper NIR surveys, like UKIDSS
(Warren, 2004) and VISTA (McPherson, Craig, \& Sutherland 2003), can
be used to go to higher extinction values. A comparison with the
current work will be important to verify that no high-extinction
patches have substantially influenced the here presented results.

\acknowledgements{This research has made use of the NASA/IPAC Infrared
Science Archive (2MASS) and the NASA/IPAC Extragalactic Database
(NED), which are operated by the Jet Propulsion Laboratory, California
Institute of Technology, under contract with the National Aeronautics
and Space Administration. The authors are grateful to the HIPASS-ZOA
team and the DENIS teams in Chile and at PDAC for all the efforts in
observing and reducing the data. We also thank Ch. Motch for making
unpublished data available to us. ACS and PAW thank the National
Science Foundation for financial support; RCKK thanks CONACyT for
their support (research grants 27602 and 40094-F).


\begin{thebibliography}{}

\bibitem {} 
Bertin, E., \& Arnouts, S. 1996, A\&A 117, 393

\bibitem {} 
Cameron, L.M. 1990, 
A\&A 233, 16

\bibitem {} 
Cardelli, J.A., Clayton, G.C., \& Mathis, J.S. 1989, 
ApJ 345, 245


\bibitem{}
Ebeling, H., Mullis, C.R., \& Tully, R.B. 2002, ApJ 580, 774

\bibitem{}
Epchtein, N. 1997, in 2nd Euroconference, The Impact of Large Scale
Near-Infrared Surveys, eds.\ F. Garz\'on et al., (Dordrecht: Kluwer), 15

\bibitem{}
Epchtein, N. 1998, in 179th Symposium of the IAU, New Horizons from
Multi-Wavelength Sky Surveys, eds.\ B.J.  McLean, D.A. Golombek,
J.J.E. Hayes, H.E. Payne, (Dordrecht: Kluwer), 106

\bibitem{}
Fairall, A.P., Woudt, P.A., \& Kraan-Korteweg, R.C. 1998, A\&ASS 127, 463


\bibitem{}
Henning, P.A., Staveley-Smith, L., Ekers, R.D., Green, A.J., et al.
2000, AJ 119, 2686 (HIZSS)


\bibitem{}
Kraan-Korteweg, R.C., Staveley-Smith, L., Donley, J., Koribalski, B.,
\& Henning, P.A. 2004, in IAU Symposium 216, Maps of the Cosmos, eds.\
M. Colless, and L. Staveley-Smith, ASP Conf.~Ser. (ASP: San
Francisco), in press, astro-ph/0311129

\bibitem{}
Kraan-Korteweg, R.C., Ochoa, M., Woudt, P., \& Andernach, H. 2004, 
these proceedings

\bibitem{}
Mamon, G.A. 1994, Ap\&SS 217, 237 (astro-ph/9312036)

\bibitem{}
Mamon, G.A. 1998, in XIVth IAP Astrophysics Meeting,
  Wide Field Surveys in Cosmology, eds.\ S. Colombi,
  Y. Mellier, \& B. Raban, (Gif-sur-Yvette: Editions Fronti\`eres), 
  323 (astro-ph/9809376)

\bibitem{}
McPherson, A.M., Craig, S.C., \& Sutherland, W. 2003, Large Ground-based
Telescopes, eds.\ J.M. Oschmann, L.M. Stepp, Proceedings of the SPIE, Volume
4837, 82


\bibitem{}
Nagayama, T., Woudt, P.A., Nagashima, C., Nakajimia, Y., et al. 
2004, MNRAS (in press)


\bibitem{}
Nagayama, T., Woudt, P.A., Nagashima, C., Nakajimia, Y., et al.
2004, these proceedings

\bibitem{}
Schr\"oder, A., Kraan-Korteweg, R.C., \& Mamon, G. 1999, PASA 16, 42

\bibitem{}
Schr\"oder, A., Mamon, G., Kraan-Korteweg, R.C., \& Woudt, P.A. 2004,
A\&A (in preparation)

\bibitem{}
Schlegel, D.J., Finkbeiner, D.P., \& Davis, M. 1998, ApJ 500, 525


\bibitem{}
Tashiro, M., Kaneda, H., Makishima, K., Iyomoto, N., et al.
1998, ApJ 499, 713

\bibitem{}
Two Micron All Sky Survey team 2003, 2MASS extended objects. Final release. vol.p.

\bibitem {} 
Visvanathan, N., \& Yamada, T. 1996, ApJS 107, 521

\bibitem{}
Warren, S. 2004, in IAU Symposium 216, Maps of the Cosmos, eds.\
M. Colless, and L. Staveley-Smith, ASP Conf.~Ser. (ASP: San
Francisco), in press

\bibitem{}
West, R.M., \& Tarenghi, M. 1989, A\&{A} 223, 61


\bibitem{}
Woudt, P.A., \&  Kraan-Korteweg, R.C., 2001, A\&{A} 380, 441

\bibitem{}
XMM-Newton Survey Science Centre (SSC) 2003: The First XMM-Newton
Serendipitous Source Catalogue, Version 1 (1XMM)

\end{thebibliography}
\end{document}